# First space and kinematical analysis of newly discovered southern UFMGs clusters with Gaia

## W. H. Elsanhoury[1, 2]

[1]Astronomy Dept., National Research Institute of Astronomy and Geophysics (NRIAG), 11421, Helwan, Cairo, Egypt (Affiliation ID: 60030681). Email: welsanhoury@gmail.com
[2]Physics Dept., Faculty of Science and Arts, Northern Border University, Rafha Branch, Saudi Arabia.

**Abstract**

The work on the kinematical parameters and spatial shape structure have been performed with Gaia DR2 astrometry data of the new recently southern discovered open clusters; UFMG 1, UFMG 2, and UFMG 3 in the vicinity (~ 1.3 degrees radius) of the rarely studied NGC 5999. The apexes positions with AD-diagram method are computed for about 107, 168, 98, and 154 members of these star clusters, respectively, our calculated values of apex coordinates, seems like: (A, D) = ($102^o.40 \pm 1.02$ & -$4^o.60 \pm 0.47$; NGC 5999), ($96^o.69 \pm 1.10$ & -$0^o.58 \pm 0.045$; UFMG 1), ($97^o.47 \pm 1.09$ & $1^o.56 \pm 0.051$; UFMG 2), and ($98^o.65 \pm 1.12$ & -$0^o.26 \pm 0.060$; UFMG 3). On the other hand, Velocity Ellipsoid Parameters VEPs for those are also computed; e.g. space velocities $\left(\overline{U}, \overline{V}, \overline{W}\right)$ due to Galactic coordinates, dispersion velocities ($\sigma_1$, $\sigma_2$, $\sigma_3$) due to matrix elements $\mu_{ij}$, projected distances ($X_\odot$, $Y_\odot$, $Z_\odot$) on the plane disc, and the Solar elements ($S_\odot$, $l_A$, $b_A$). According to an approximation of spatial and kinematical shape, UFMGs and NGC 5999 seem to have a spatial difference in their space locations but they appear to have formed in the same region of the Galactic disc. The total cumulative mass $M_C$; including total number of main-sequence $N_{MS}$ and non-main-sequence $N_{non-MS}$ of these clusters also evaluated here with a second-order polynomial of mass-luminosity relation in order to get clusters tidal radii (pc). Finally, we concluded that NGC 5999, UFMG 1, and UFMG 2 are dynamically relaxed (i.e. $\tau \gg 1$), and the fourth one in non-relaxed.

## 1. Introduction

Open clusters are ensembles of Gravitationally bound stars moving with almost identical space motions (Montes et al. 2001; Chumak & Rastorguev 2006). These objects have long been recognized as important tools in the study of the Galactic disc, also provides a key insight into the scheme of stellar evolution. Old and intermediate-age open clusters are excellent probes of early disc evolution, while young open clusters provide information about current star formation processes and are key objects to delineate Galactic structure (Piatti et al., 1999).

Galactic open clusters have in general, large angular sizes and low concentrations of stars. Clusters of small angular size are also fundamental as calibrators of integrated properties in view of population synthesis studies in Galaxies (see, e.g., Bica & Alloin 1986 & 1987) and the determination of astrophysical parameters of star clusters in distant Galaxies for which it is not possible to obtain color-magnitude diagrams (CMDs).

Piatti et al. (1999) investigate in-depth NGC 5999 open cluster beside the search for variable stars within the OGLE survey (Pietrzynski et al. 1998). NGC 5999 is an interesting target which represents an open cluster as well as a stellar moving group.

NGC 5999 (other name C1548−563) located in the direction of southern disc with $\alpha_{2000} = 15^{hh} 52^{mm} 11^{ss}.30$ and $\delta_{2000} = -56^{dd} 29^{mm} 17^{ss}.00$ and Galactic coordinates $l = 326^{o}.00$ and $b = -1^{o}.94$ (Ferreira et al. 2019), 400 Myr old, reddening $E(B − V) = 0.45 ± 0.05$, and distance ranges from 1.6 to 2.5 kpc (Dias et al. 2002; Piatti et al. 1999; Kharchenko et al. 2013; Netopil et al. 2007; Santos & Bica 1993; Moni Bidin et al. 2014).

Based on integrated spectra and template matching, Santos & Bica (1993) founded its $T_{age}$ (log) = 8.0 yr, additionally they also obtained a slightly higher age of ∼230 Myr using Balmer lines and metallicity of [Fe/H] = 0.18 from Ca II triplet lines. Netopil et al. (2007) assumed Solar metallicity (Z = 0.019) for the isochrones fitting.

Kharachenko et al. (2013) with Milky Way Star Clusters project MWSC gives log ($T_{age}$) = 8.6 ± 0.095 yr, reddening about 0.437, distance 1629 pc, and other some internal structure parameters like; core and tidal radii are about 0.71 ± 0.09 pc and 6.05 ± 0.83 pc respectively. The projected distances ($X_{\odot}$, $Y_{\odot}$, $Z_{\odot}$) pc and space velocities (U, V, W) km/s have been founded with their uncertainties by Soubiran et al. (2018)[1] and Gaia DR2[2] (Gaia Collaboration et al. 2018), also radial velocity of NGC 5999 is of the order of -30.78 ± 0.48 km/s (Soubiran et al. 2018).

For a spatial distribution in VPD of the region (radius ≃ 1.3 degrees) adjacent the intermediate-age of NGC 5999 (immersed in a dense star field), Ferreira et al. (2019)

---

[1] http://vizier.cfa.harvard.edu/viz-bin/VizieR?-source=J/A+A/619/A155
[2] http://vizier.u-strasbg.fr/viz-bin/VizieR-3?-source=I/345&-out.add=_r

discovered existence new three open clusters not reported before; i.e. UFMG 1, UFMG 2, and UFMG 3 by using Gaia DR2 data. The abbreviation of UFMG is due to Universidade Federal de Minas. Gerais.

Gaia Collaboration et al. (2018) and Evans et al. (2018), provides positions, proper motions (ranged from 0.06 to 1.20 mas/yr) in both directions, parallaxes (from 0.04 to 0.70 mas) and magnitudes in three passbands of the Gaia photometric systems; i.e. G, $G_{BP}$ and $G_{RP}$ with precision at the mmag level, for about 1.3 billion sources, as well as kinematics to a large number of cluster members with an accuracy not available before. This newly available parameter space promises to open a new window on cluster disruption and the build-up of the field population (Fürnkranz et al. 2019).

The aimed of the paper is to scope on kinematics, luminosity and mass functions, and dynamics of rarely studied NGC 5999 cluster, and first structure analyses of these three new discovered UFMGs open clusters, the analyses were done for membership selected, provides their elliptical motions, projected distances, space and dispersion velocities, total masses, velocity ellipsoid parameters VEPs … etc.

The structure of this paper is organized as follows: Section 2 describes the observational data with Gaia DR2 catalog. Section 3 is devoted to the kinematical structure. Section 4 deals with the luminosity and mass functions. In Section 5, we discuss the dynamical relaxation time. The conclusion of the current work is presented in Section 6.

## 2. Data with the Gaia DR2
### 2.1 Data gathering

Ferreira et al. (2019) present the spatial distribution of NGC 5999 and these three newly discovered open clusters, namely UFMG 1, UFMG 2 and UFMG 3 as shown in Fig. 1.

On the other hand, Table 1 presents the basic (fundamental) data of NGC 5999 and those objects devoted with Ferreira et al. (2019) and the Milky Way Star Clusters project (MWSC; Kharchenko et al. 2013) which is based on 2MASS (Skrutskie et al. 2006) photometry and PPMXL (Roeser et al. 2010) astrometry.

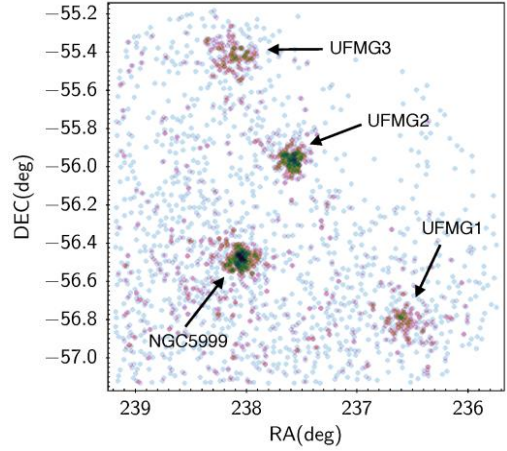

**Fig. 1:** The spatial distribution of NGC5999 and the three discovered clusters.

**Table 1:** The fundamental parameters of open clusters NGC 5999, UFMG 1, UFMG 2, and UFMG 3.

|  | **NGC 5999** | **UFMG 1** | **UFMG 2** | **UFMG 3** | **Ref.** |
|---|---|---|---|---|---|
| $\alpha$ | $15^{hh}\,52^{mm}\,11^{ss}.30$ | $15^{hh}\,46^{mm}\,24^{ss}.50$ | $15^{hh}\,50^{mm}\,23^{ss}.30$ | $15^{hh}\,52^{mm}\,26^{ss}.20$ | Ferreira et al. (2019) |
|  | $15^{hh}\,52^{mm}\,10^{ss}.20$ | - | - | - | Kharchenko et al. (2013) |
| $\delta$ | $-56^{dd}\,29^{mm}\,17^{ss}.00$ | $-56^{dd}\,48^{mm}\,29^{ss}.00$ | $-55^{dd}\,57^{mm}\,32^{ss}.00$ | $-55^{dd}\,25^{mm}\,19^{ss}.00$ | Ferreira et al. (2019) |
|  | $-56^{dd}\,28^{mm}\,01^{ss}.20$ | - | - | - | Kharchenko et al. (2013) |
| $l$ | $326°.00$ | $325°.19$ | $326°.14$ | $326°.70$ | Ferreira et al. (2019) |
|  | $326°.02$ | - | - | - | Kharchenko et al. (2013) |
| $b$ | $-1°.94$ | $-1°.70$ | $-1°.37$ | $-1°.13$ | Ferreira et al. (2019) |
|  | $-1°.92$ | - | - | - | Kharchenko et al. (2013) |
| $r_{lim}$ (arcmin) | $8.70 \pm 1.13$ | $13.71 \pm 1.52$ | $11.15 \pm 1.39$ | $13.66 \pm 1.59$ | Ferreira et al. (2019) |
|  | 3.3 | - | - | - | Kharchenko et al. (2013) |
| $r_c$ (arcmin) | $2.83 \pm 0.38$ | $4.13 \pm 0.44$ | $3.72 \pm 0.46$ | $5.69 \pm 0.91$ | Ferreira et al. (2019) |
|  | $1.50 \pm 0.18$ | - | - | - | Kharchenko et al. (2013) |
| $r_t$ (arcmin) | $11.33 \pm 1.32$ | $22.63 \pm 3.26$ | $14.87 \pm 1.63$ | $16.62 \pm 2.28$ | Ferreira et al. (2019) |
|  | $12.77 \pm 1.75$ | - | - | - | Kharchenko et al. (2013) |
| $\log(T_{age})$ (yr) | $8.50 \pm 0.10$ | $8.90 \pm 0.05$ | $9.15 \pm 0.05$ | $8.00 \pm 0.10$ | Ferreira et al. (2019) |
|  | $8.60 \pm 0.095$ | - | - | - | Kharchenko et al. (2013) |
| d (kpc) | $1.82 \pm 0.19$ | $1.58 \pm 0.16$ | $1.48 \pm 0.15$ | $1.51 \pm 0.14$ | Ferreira et al. (2019) |
|  | 1.629 | - | - | - | Kharchenko et al. (2013) |
| Z | 0.009 | 0.009 | 0.015 | 0.015 | Ferreira et al. (2019) |

In what follows, and in order to get a net worksheet data (members) with Gaia Collaboration et al. (2018) for these four open clusters under study, we restrict our candidates for the following; i) we use the criteria devoted with Roeser et al. (2010) into which the stars with proper motion in both directions ($\mu_\alpha \cos\delta$ & $\mu_\delta$) uncertainties $\geq 4.0$ mas yr$^{-1}$ have been removed, ii) stars with an observational uncertainties $\geq 0.20$ mag have been omitted (Claria & Lapasset 1986), iii) our calculations discover stars for which the parallaxes $\pi$ (mas) are in negative values, therefore we neglect six stars with NGC 5999 having Gaia DR2 ID sources; 5836054686390674048, 5836057125932361088, 5836057877418617088, 5884000913721642752, 5884094578368452096, and 5884100389516500736; zero stars for UFMG 1; four stars with UFMG 2 having IDs; 5884123681059862272, 5884173129091382656,

5884175121955857536, and 5884177011741669760; and three stars with UFMG 3 having IDs 5884162885531807488, 5884541877779375744, and 5884548337410698368, and iv) on the same manner, we omitted about 19, 42, 19, and 27 stars of these open clusters which having radial velocities significantly exceeding than the average values, this fact led to a significant deviation of the direction of apex of these stars. Finally, after excluding these above-mentioned stars, we considered in our calculation about 107, 168, 98, and 154 as highly astrometric probable candidates for these four open clusters respectively. The obtained results are given in Table 2 with the following format:

Column 1: The cluster name and its member counts (N).

Column 2: The Gaia DR2 source ID number.

Columns 3 and 4: The coordinate elements, i.e. right ascension and declination (degrees).

Columns 5 and 6: The parallax and its uncertainties (mas).

Columns 7 and 8: The proper motion along right ascension and its uncertainties (mas/yr).

Columns 9 and 10: The proper motion along with declination and its uncertainties (mas/yr).

Columns 11 and 12: The radial velocity and its uncertainties (km/s).

**Table 2:** Data of star members for NGC 5999, UFMG 1, UFMG 2, and UFMG 3 open clusters taken from Gaia DR2 catalog.

| Cluster | Gaia DR2 Source | α deg | δ deg | π mas | $\sigma_\pi$ mas | $\mu_\alpha \cos\delta$ mas/yr | $\sigma_{\mu\alpha}$ mas/yr | $\mu_\delta$ mas/yr | $\sigma_{\mu\delta}$ mas/yr | $V_r$ km/s | $\sigma_{Vr}$ km/s |
|---|---|---|---|---|---|---|---|---|---|---|---|
| **NGC 5999 (107 members)** | | | | | | | | | | | |
| 1 | 5836054755110030080 | 238.1735 | -56.6204 | 0.3774 | 0.0863 | -6.293 | 0.117 | -6 | 0.112 | -23.78 | 0.27 |
| 2 | 5836055133067276416 | 238.093 | -56.6281 | 0.1212 | 0.0554 | -4.009 | 0.08 | -5.621 | 0.074 | -103.46 | 0.96 |
| 3 | 5836055064347804800 | 238.1085 | -56.6335 | 0.1718 | 0.0804 | -3.274 | 0.116 | -4.354 | 0.108 | -32.07 | 0.87 |
| . | . | . | . | . | . | . | . | . | . | . | . |
| . | . | . | . | . | . | . | . | . | . | . | . |
| . | . | . | . | . | . | . | . | . | . | . | . |
| 107 | 5884103928570801792 | 238.0492 | -56.3487 | 0.3758 | 0.0444 | -4.574 | 0.057 | -6.778 | 0.054 | -46.03 | 0.44 |
| **UFMG 1 (168 members)** | | | | | | | | | | | |
| 1 | 5882522070586168192 | 236.6126 | -57.0105 | 0.1514 | 0.0433 | -6.59 | 0.05 | -5.165 | 0.052 | -41.38 | 2.06 |
| 2 | 5882522349837650688 | 236.5727 | -56.9939 | 0.2748 | 0.0371 | -2.943 | 0.046 | -2.616 | 0.045 | -22.89 | 5.94 |
| 3 | 5882523101379067392 | 236.4795 | -56.982 | 0.6321 | 0.0555 | -4.799 | 0.07 | -10.159 | 0.067 | -38.04 | 0.11 |
| . | . | . | . | . | . | . | . | . | . | . | . |
| . | . | . | . | . | . | . | . | . | . | . | . |
| . | . | . | . | . | . | . | . | . | . | . | . |
| 168 | 5884044073932898944 | 236.4449 | -56.6096 | 0.635 | 0.0372 | 2.374 | 0.051 | -3.68 | 0.049 | -24.78 | 0.49 |
| **UFMG 2 (98 members)** | | | | | | | | | | | |
| 1 | 5884118875058684416 | 237.6082 | -56.1412 | 0.7455 | 0.0355 | -7.662 | 0.05 | -4.281 | 0.045 | -32.35 | 0.53 |
| 2 | 5884119459174245888 | 237.4951 | -56.1255 | 0.5227 | 0.0861 | -4.363 | 0.127 | -4.511 | 0.117 | -75.16 | 0.57 |
| 3 | 5884119592250900736 | 237.568 | -56.1237 | 0.7617 | 0.0796 | -3.211 | 0.102 | -4.365 | 0.103 | -34.95 | 0.2 |
| . | . | . | . | . | . | . | . | . | . | . | . |
| . | . | . | . | . | . | . | . | . | . | . | . |
| . | . | . | . | . | . | . | . | . | . | . | . |
| 98 | 5884200342005310848 | 237.6462 | -55.8079 | 1.5129 | 0.0366 | -8.856 | 0.052 | -6.229 | 0.045 | -42.28 | 2.41 |
| **UFMG 3 (154 members)** | | | | | | | | | | | |
| 1 | 5884159453914021120 | 238.1917 | -55.6197 | 0.5114 | 0.023 | -1.125 | 0.036 | -3.5 | 0.031 | -6.01 | 1.31 |
| 2 | 5884159385194497024 | 238.2322 | -55.6326 | 0.2915 | 0.0417 | -2.708 | 0.059 | -2.367 | 0.053 | -57.44 | 5.11 |
| 3 | 5884159453914026368 | 238.1957 | -55.6154 | 0.3481 | 0.0398 | 2.823 | 0.058 | -3.756 | 0.049 | -45.12 | 1.18 |
| . | . | . | . | . | . | . | . | . | . | . | . |
| . | . | . | . | . | . | . | . | . | . | . | . |
| . | . | . | . | . | . | . | . | . | . | . | . |
| 154 | 5884596441039489024 | 238.1175 | -55.1959 | 0.4271 | 0.046 | -6.844 | 0.07 | -4.298 | 0.063 | -18.17 | 0.73 |

## 3. The kinematical structure

In what follows, we need to study the kinematical processes takes place in/out of NGC 5999 with those recently discovered open clusters. Our obtained results will appear here in Table 3.

Let us drive the basic equations required for our kinematical analysis. Consider d is the distance of the star members from the Sun along with the equatorial system, and because of the observational data for proper motions ($\mu_\alpha \cos\delta$ & $\mu_\delta$) are always given in the equatorial motion, then the position and velocities given here such that (Mihalas & Binney, 1981), i.e.

$$x = d \cos\delta \cos\alpha, \quad (1)$$

$$y = d \cos\delta \sin\alpha, \quad (2)$$

$$z = d \sin\delta. \tag{3}$$

Differentiating Equations (1), (2), and (3) with respect to time, we have the equatorial components of stellar velocity along (x, y, z) axes with respect to Sun. i.e.

$$V_x = -4.74 d\, \mu_\alpha \cos\delta \sin\alpha - 4.74 d\, \mu_\delta \sin\delta \cos\alpha + \dot{d} \cos\delta \cos\alpha, \tag{4}$$

$$V_y = +4.74 d\, \mu_\alpha \cos\delta \cos\alpha - 4.74 d\, \mu_\delta \sin\delta \sin\alpha + \dot{d} \cos\delta \sin\alpha, \tag{5}$$

$$V_z = +4.74 d\, \mu_\delta \cos\delta + \dot{d} \sin\delta. \tag{6}$$

The quantity $\dot{d}$ stands for the radial velocity $V_r$ (km / s).

- *Vertex (apex position) of the cluster*

Considering the cluster members moves with equal space velocities, the equatorial coordinates, and the proper motion components of the cluster stars are known. In what follows, we need to determine the equatorial coordinates of the vertex or apex (sometimes called the convergent point), that is the right ascension and declination (A, D) of the cluster members, the determination of (A, D) is one of the most important problems in the kinematical and physical studies of moving clusters (Wayman 1965; Hanson 1975; Eggen 1984; Gunn et al. 1988).

Along with the classical methods of research, we will put in practice a unified approach used in the study of the Ursa Major kinematic stream Chupina et al. (2001 and 2006) and many other objects. It based on the method of apex chart (AD-diagram). This method has been used by our group to determine apex and other kinematical parameters for open clusters M67, NGC 188 and Pleiades (Vereshchagin et al. 2014; Elsanhoury et al. 2016 & 2018; Elsanhoury & Nouh 2019). AD-chart (or AD-diagram) represents the distribution of individual stars apexes positions in the equatorial coordinate system and could be used to study the kinematical structure of the clusters and to find out its inner kinematical substructures.

The coordinates obtained by solving the geometric problems in which the intersection of vectors of spatial velocities (i.e. $V_x$, $V_y$, $V_z$) of stars with the celestial sphere, with the beginning of the vectors, moved to the point of observations. A formal description of the method, diagramming technique, and formulas determining the error ellipses can be found in Chupina et al. (2001).

Vereshchagin et al. (2014), and Elsanhoury et al. (2016) used the method of apex diagram (AD-diagram) as given by Chupina et al. (2001 and 2006), into which the equatorial coordinates of the convergent point having the following forms:

i.e.

$$A = \tan^{-1}\left(\overline{V_y}/\overline{V_x}\right), \tag{7}$$

$$D = \tan^{-1}\left(\overline{V_z}/\sqrt{\overline{V_x}^2 + \overline{V_y}^2}\right). \tag{8}$$

The apex equatorial coordinates for NGC 5999 and these three UFMGs open clusters are shown in Fig. 2.

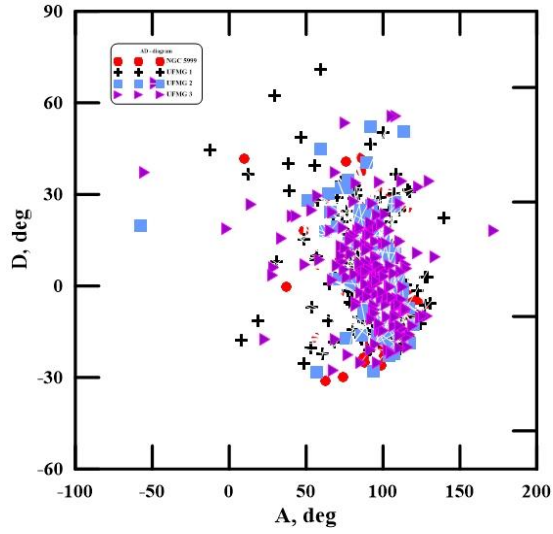

**Fig. 2:** The AD-diagram for NGC 5999 (solid closed red 107 circles), UFMG 1 (solid closed black 168 plus symbol), UFMG 2(solid closed blue 98 squares), and UFMG 3 (solid closed purple 154 triangles).

- *The components of space velocity*

Now we compute the space velocity components U, V and W along the three axes with Liu et al. (2011) which used an equatorial Galactic transformation. They determined the position of the Galactic plane using recent catalogs like Two-Micron All-Sky Survey (2MASS, Skrutskie et al. 2006) and defined the optimal Galactic coordinate system by adopting the ICRS position of the compact radio source Sagittarius A$^*$.

i.e.

$$U = -0.0518807421 V_x - 0.8722226427 V_y - 0.4863497200 V_z, \tag{9}$$

$$V = 0.4846922369 V_x - 0.4477920852 V_y + 0.7513692061 V_z, \tag{10}$$

$$W = -0.8731447899 V_x - 0.1967483417 V_y + 0.4459913295 V_z. \tag{11}$$

While the mean velocities are given by:

$$\bar{U} = \frac{1}{N}\sum_{i=1}^{N} U_i, \qquad (12)$$

$$\bar{V} = \frac{1}{N}\sum_{i=1}^{N} V_i, \qquad (13)$$

$$\bar{W} = \frac{1}{N}\sum_{i=1}^{N} W_i. \qquad (14)$$

Fig. 2 shows the vector point diagram VPD, into which the components U, V, and W of the spatial velocities of the cluster member stars sketched such that; (U *vs.* V) and (W *vs.* V). As can be seen from Fig. 3, the components of space velocities of all member stars are close. This devotes that the clusters are flow moves in approximately the same direction.

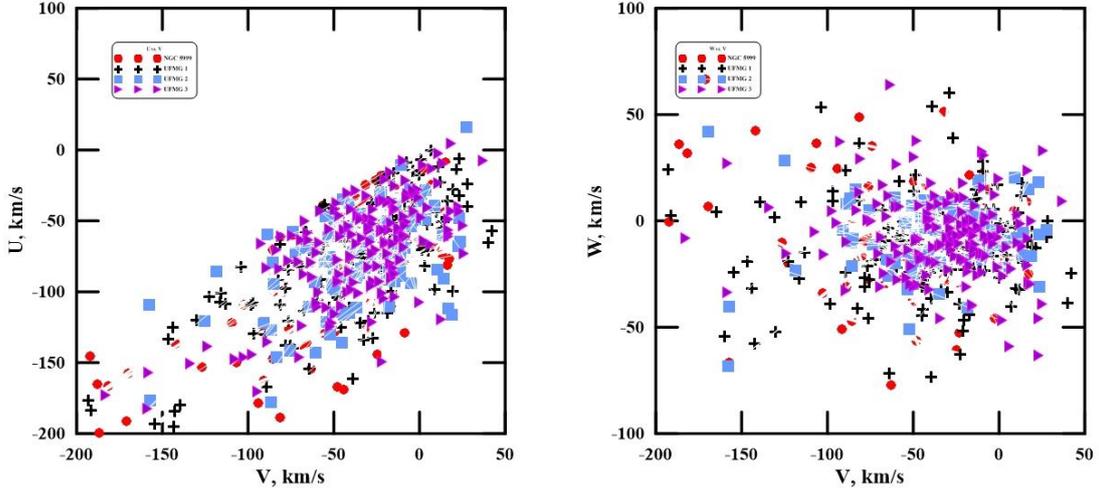

**Fig. 3:** Comparison of the spatial velocities of the member stars of the clusters. The U (km s$^{-1}$) axis is directed to the anti-center of the Galaxy, the V (km s$^{-1}$) axis is in the direction of rotation of the Galaxy, and the W (km s$^{-1}$) axis is toward the north pole of the Galaxy. Cluster stars from the list are shown, NGC 5999 (solid closed red 107 circles), UFMG 1 (solid closed black 168 plus symbol), UFMG 2 (solid closed blue 98 squares), and UFMG 3 (solid closed purple 154 triangles).

- *The distance r of the cluster*

Since, the distance r (pc) is the reciprocal of the parallax p, then the distance of the cluster have the following form

$$r = N \Big/ \sum_{i=1}^{N} p_i . \qquad (15)$$

- *The center ($x_c$, $y_c$, $z_c$) of the cluster*

The center of the cluster can be derived by the simple method of finding the equatorial coordinates of the center of mass for the number $N_i$ of discrete objects, i.e.

$$x_c = \left[\sum_{i=1}^{N} d_i \cos\alpha_i \cos\delta_i\right] / N, \qquad (16)$$

$$y_c = \left[\sum_{i=1}^{N} d_i \sin\alpha_i \cos\delta_i\right] / N, \qquad (17)$$

$$z_c = \left[\sum_{i=1}^{N} d_i \sin\delta_i\right] / N. \qquad (18)$$

- *The velocity ellipsoid parameters (VEPs)*

To compute the parameters of the ellipsoidal motion and its parameters for NGC 5999, UFMG 1, UFMG 2, and UFMG 3 clusters, we used a computational algorithm presented by Elsanhoury et al. (2013 & 2015). A brief explanation of the algorithm will be given here. The coordinates of the i$^{th.}$ Star with respect to axes parallel to the original axes, but shifted to the center of the distribution, i.e. to the point $\bar{U}, \bar{V}$, and $\bar{W}$, will be $(U_i - \bar{U}); (V_i - \bar{V}); (W_i - \bar{W})$.

Let ξ be an arbitrary axis, its zero point coincident with the center of the distribution and let *l,m* and *n* be the direction cosines of the axis with respect to the shifted ones; then the coordinates $Q_i$ of the point *i*, with respect to the ξ-axis are given by:

$$Q_i = l(U_i - \bar{U}) + m(V_i - \bar{V}) + n(W_i - \bar{W}). \qquad (19)$$

Considering $\sigma^2$ is a generalization of the mean square deviation, i.e.

$$\sigma^2 = \frac{1}{N}\sum_{i=1}^{N} Q_i^2 \qquad (20)$$

Due to mean velocities $\bar{U}, \bar{V}$, and $\bar{W}$, and Equations (19) & (20) we deduce after some calculations that

$$\sigma^2 = \underline{x}^T B \underline{x} \qquad (21)$$

where $\underline{x}$ is the $(3 \times 1)$ direction cosines vector, and $B$ is $(3 \times 3)$ symmetric matrix elements $\mu_{ij}$

and

$$\mu_{11} = \frac{1}{N}\sum_{i=1}^{N} U_i^2 - \left(\overline{U}\right)^2; \quad \mu_{12} = \frac{1}{N}\sum_{i=1}^{N} U_i V_i - \overline{UV};$$

$$\mu_{13} = \frac{1}{N}\sum_{i=1}^{N} U_i W_i - \overline{UW}; \quad \mu_{22} = \frac{1}{N}\sum_{i=1}^{N} V_i^2 - \left(\overline{V}\right)^2; \quad (22)$$

$$\mu_{23} = \frac{1}{N}\sum_{i=1}^{N} V_i W_i - \overline{VW}; \quad \mu_{33} = \frac{1}{N}\sum_{i=1}^{N} W_i^2 - \left(\overline{W}\right)^2.$$

Now, the necessary conditions for an extremum are

$$(B - \lambda I)\underline{x} = 0 \quad (23)$$

These are three homogenous equations in three unknowns have a nontrivial solution if and only if

$$D(\lambda) = |B - \lambda I| = 0, \quad (24)$$

Equation (24) is the characteristic equation for the matrix $B$, where $\lambda$ is the eigenvalue, $\underline{x}$ and $B$ could be written as

$$\underline{x} = \begin{bmatrix} l \\ m \\ n \end{bmatrix} \text{ and } B = \begin{vmatrix} \mu_{11} & \mu_{12} & \mu_{13} \\ \mu_{12} & \mu_{22} & \mu_{23} \\ \mu_{13} & \mu_{23} & \mu_{33} \end{vmatrix}$$

Depending on the matrix that controls the eigenvalue problem [Equation (23)] for the velocity ellipsoid, we established analytical expressions of some parameters in terms of the matrix elements $\mu_{ij}$.

- ***The $\sigma_i$; $i$ =1, 2, 3 parameters***

The $\sigma_i$; $i$ =1, 2, 3 parameters are defined as the square roots of the eigenvalues $\lambda_i$, i.e.

$$\sigma_i = \sqrt{\lambda_i} \quad (25)$$

where

$$\lambda_1 = 2\rho^{\frac{1}{3}} \cos\frac{\phi}{3} - \frac{k_1}{3};$$

$$\lambda_2 = -\rho^{\frac{1}{3}} \left\{\cos\frac{\phi}{3} + \sqrt{3}\sin\frac{\phi}{3}\right\} - \frac{k_1}{3}; \quad (26)$$

$$\lambda_3 = -\rho^{\frac{1}{3}} \left\{\cos\frac{\phi}{3} - \sqrt{3}\sin\frac{\phi}{3}\right\} - \frac{k_1}{3}.$$

where

$$k_1 = -(\mu_{11} + \mu_{22} + \mu_{33}),$$
$$k_2 = \mu_{11}\mu_{22} + \mu_{11}\mu_{33} + \mu_{22}\mu_{33} - (\mu_{12}^2 + \mu_{13}^2 + \mu_{23}^2), \quad (27)$$
$$k_3 = \mu_{12}^2\mu_{33} + \mu_{13}^2\mu_{22} + \mu_{23}^2\mu_{11} - \mu_{11}\mu_{22}\mu_{33} - 2\mu_{12}\mu_{13}\mu_{23}.$$

$$q = \frac{1}{3}k_2 - \frac{1}{9}k_1^2 \quad ; \quad r = \frac{1}{6}(k_1 k_2 - 3k_3) - \frac{1}{27}k_1^3 \quad (28)$$

$$\rho = \sqrt{-q^3} \quad (29)$$

$$x = \rho^2 - r^2 \quad (30)$$

and

$$\phi = \tan^{-1}\left(\frac{\sqrt{x}}{r}\right) \quad (31)$$

- ***The E parameter***

This represents the volume of the ellipsoid, i.e.

$$E = \frac{4}{3}\pi\sigma_1\sigma_2\sigma_3. \quad (32)$$

- ***The $l_i$, $m_i$, and $n_i$ parameters***

The direction cosines $l$, $m$, and $n$ ($\forall\ i = 1, 2, 3$) are mathematically represented as follows:

$$l_i = \left[\mu_{22}\mu_{33} - \sigma_i^2(\mu_{22} + \mu_{33} - \sigma_i^2) - \mu_{23}^2\right]/D_i, \quad (33)$$

$$m_i = \left[\mu_{23}\mu_{13} - \mu_{12}\mu_{33} + \sigma_i^2\mu_{12}\right]/D_i, \quad (34)$$

$$n_i = \left[\mu_{12}\mu_{23} - \mu_{13}\mu_{22} + \sigma_i^2\mu_{13}\right]/D_i. \quad (35)$$

where

$$\begin{aligned} D_i^2 &= (\mu_{22}\mu_{33} - \mu_{23}^2)^2 + (\mu_{23}\mu_{13} - \mu_{12}\mu_{33})^2 + (\mu_{12}\mu_{23} - \mu_{13}\mu_{22})^2 \\ &\quad + 2\left[(\mu_{22} + \mu_{33})(\mu_{23}^2 - \mu_{22}\mu_{33}) + \mu_{12}(\mu_{23}\mu_{13} - \mu_{12}\mu_{33}) + \mu_{13}(\mu_{12}\mu_{23} - \mu_{13}\mu_{22})\right]\sigma_i^2 \\ &\quad + (\mu_{33}^2 + 4\mu_{22}\mu_{33} + \mu_{22}^2 - 2\mu_{23}^2 + \mu_{12}^2 + \mu_{13}^2)\sigma_i^4 - 2(\mu_{22} + \mu_{33})\sigma_i^6 + \sigma_i^8. \end{aligned}$$

- ***Projected distances***

Due to our estimated heliocentric distances r, we can infer the cluster's distance to the Galactic center $R_{gc}$ (Mihalas & Binney, 1981) as a function of the Sun's distance from the Galactic center $R_o = 7.5 \pm 0.3$ kpc (Francis & Anderson, 2014), i.e. $R_{gc}^2 = R_o^2 + r^2 - 2R_o r \cos l$, in such a way the projected distances to the Galactic plane $X_\odot$ and $Y_\odot$, and the distance from the Galactic plane $Z_\odot$ (Tadross 2011) could computed as follows:

$$X_\odot = r\cos(b)\cos(l),\ Y_\odot = r\cos(b)\sin(l),\ \text{and}\ Z_\odot = r\sin(b). \tag{36}$$

- *The Solar elements*

The Solar motion can be defined as the absolute value of the Sun's velocity relative to the group of stars under consideration,

i.e.

$$S_\odot = \left(\overline{U}^2 + \overline{V}^2 + \overline{W}^2\right)^{1/2}\ km\ s^{-1}, \tag{37}$$

The Galactic longitude $(l_A)$ and Galactic latitude $(b_A)$ of the Solar apex are

$$l_A = tan^{-1}\left(-\overline{V}/\overline{U}\right), \tag{38}$$

$$b_A = sin^{-1}\left(-\overline{W}/S_\odot\right). \tag{39}$$

- *Shape in space*

Figure 4 shows the 3D distribution of member stars of these open clusters under investigation. We can notice that the member stars stretched at an angle of approximately 45° relative to the axis OX, i.e. reveals us to indicate those members elongated along the Galactic center, and located in separate regions in space.

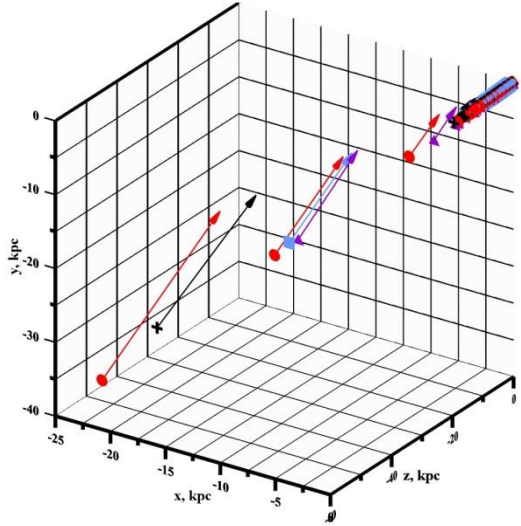

**Fig. 4:** The 3D distribution of member stars of open clusters under consideration, NGC 5999 (solid closed red 107 circles), UFMG 1 (solid closed black 168 plus symbol), UFMG 2 (solid closed blue 98 squares), and UFMG 3 (solid closed purple 154 triangles).

**Table 3:** Our kinematical parameters of NGC 5999, UFMG 1, UFMG 2, and UFMG 3 with other published ones.

| Parameter | NGC 5999 | UFMG 1 | UFMG 2 | UFMG 3 | Ref. |
|---|---|---|---|---|---|
| No. of members (N) | 107 | 168 | 98 | 154 | Present work |
|  | 405 | 191 | 592 | 261 | Ferreira et al. (2019) |
| (A, D) | $102°.40 \pm 1.02, -4°.60 \pm 0.47$ | $96°.69 \pm 1.10, -0°.58 \pm 0.045$ | $97°.47 \pm 1.09, 1°.56 \pm 0.051$ | $98°.65 \pm 1.12, -0°.26 \pm 0.060$ | Present work |
| r (kpc) | $2.16 \pm 0.046$ | $1.77 \pm 0.042$ | $1.96 \pm 0.044$ | $1.55 \pm 0.039$ | Present work |
|  | $1.82 \pm 0.19$ | $1.58 \pm 0.16$ | $1.48 \pm 0.15$ | $1.51 \pm 0.14$ | Ferreira et al. (2019) |
|  | 1.629 | - | - | - | Kharchenko et al. (2013) |
| $\overline{V}_x$ (km s$^{-1}$) | -29.76 | -11.61 | -13.31 | -13.95 | Present work |
| $\overline{V}_y$ (km s$^{-1}$) | 135.44 | 99.01 | 101.53 | 91.68 | Present work |
| $\overline{V}_z$ (km s$^{-1}$) | -11.17 | -1.00 | 2.79 | 0.42 | Present work |
| $\overline{U}$ (km s$^{-1}$) | -111.16 | -85.27 | -89.23 | -79.45 | Present work |
|  | $-65.37 \pm 0.52$ | - | - | - | Soubiran et al. (2018) |
| $\overline{V}$ (km s$^{-1}$) | -83.46 | -50.71 | -49.82 | -47.50 | Present work |
|  | $-41.01 \pm 0.56$ | - | - | - | Soubiran et al. (2018) |
| $\overline{W}$ (km s$^{-1}$) | -5.65 | -9.80 | -7.11 | -5.68 | Present work |
|  | $-13.98 \pm 0.16$ | - | - | - | Soubiran et al. (2018) |
| Space vel. | 139.12 | 99.69 | 102.44 | 92.74 | Present work |
|  | $78.43 \pm 1.51$ | - | - | - | Soubiran et al. (2018) |
| $(\lambda_1, \lambda_2, \lambda_3)$ (km s$^{-1}$) | 84853.9, 2464.64, 746.33 | 30723.5, 2530.15, 625.13 | 28226.0, 2024.47, 289.64 | 20611.7, 1412.98, 687.39 | Present work |
| $(\sigma_1, \sigma_2, \sigma_3)$ (km s$^{-1}$) | 291.30, 49.65, 27.32 | 175.28, 50.31, 25.03 | 168.01, 45.00, 17.02 | 143.57, 37.59, 26.22 | Present work |
| $\sigma$ (km s$^{-1}$) | $296.76 \pm 17.23$ | $184.06 \pm 13.57$ | $174.76 \pm 13.22$ | $150.71 \pm 12.28$ | Present work |
| E (kpc$^3$) | $1655 \pm 41$ | $925 \pm 30$ | $540 \pm 23$ | $593 \pm 24$ | Present work |
| $l_1, m_1, n_1$ (degree) | 0.63, 0.76, -0.06 | 0.68, 0.73, -0.11 | 0.68, 0.73, 0.08 | 0.71, 0.70, 0.03 | Present work |
| $l_2, m_2, n_2$ (degree) | -0.74, 0.58, -0.34 | -0.57, 0.43, -0.70 | -0.73, 0.68, 0.09 | -0.70, 0.71, -0.06 | Present work |
| $l_3, m_3, n_3$ (degree) | 0.23, -0.26, -0.94 | 0.46, -0.54, -0.71 | 0.01, -0.13, 1.00 | 0.06, -0.03, -1.00 | Present work |
| $x_c, y_c, z_c$ (kpc) | -1.31, -2.10, -3.74 | -1.04, -1.58, -2.89 | -0.904, -1.43, -2.501 | -0.874, -1.41, -2.40 | Present work |
| $X_\odot$ (kpc) | 1.79 | 1.45 | 1.63 | 1.30 | Present work |
|  | 2.33 | - | - | - | Cantat-Gaudin et al. (2018) |
|  | $-6.00 \pm 0.019$ | - | - | - | Soubiran et al. (2018) |
| $Y_\odot$ (kpc) | -1.21 | -1.01 | -1.10 | -0.85 | Present work |
|  | -1.57 | - | - | - | Cantat-Gaudin et al. (2018) |
|  | $-1.57 \pm 0.013$ | - | - | - | Soubiran et al. (2018) |
| $Z_\odot$ (kpc) | -0.07 | -0.05 | -0.05 | -0.04 | Present work |
|  | -0.095 | - | - | - | Cantat-Gaudin et al. (2018) |
|  | $-0.08 \pm 0.0001$ | - | - | - | Soubiran et al. (2018) |
| $R_{gc}$ (kpc) | 5.84 | 6.13 | 5.97 | 6.27 | Present work |
| $S_\odot$ (km s$^{-1}$) | 139.12 | 99.69 | 102.44 | 92.74 | Present work |
| $l_A$ | $-36°.90$ | $-30°.74$ | $-29°.18$ | $-30°.88$ | Present work |
| $b_A$ | $2°.33$ | $5°.64$ | $3°.98$ | $3°.51$ | Present work |

## 4. Luminosity function and mass function

In what follows, we need to determine both luminosity and mass functions (LF & MF) for 107, 168, 98, and 154 member stars of NGC 5999, UFMG 1, UFMG 2, and UFMG 3 open clusters respectively.

Let us now turn to determine the LF, which is one of the important intrinsic properties of stars in clusters. Fig. 5 presents the LF of these four open clusters, where incompleteness of the LF is present for absolute G magnitudes in the range $-5 < M_G < 6$ for NGC 5999; $-4 < M_G < 8$ for UFMG 1; $-5 < M_G < 5$ for UFMG 2; and $-5 < M_G < 7$ for UFMG 3, this incompleteness of the photometry affects the shape of the luminosity function of these clusters. From LFs of these open clusters, we can conclude that the massive bright stars seem to be centrally concentrated more than the lower ones.

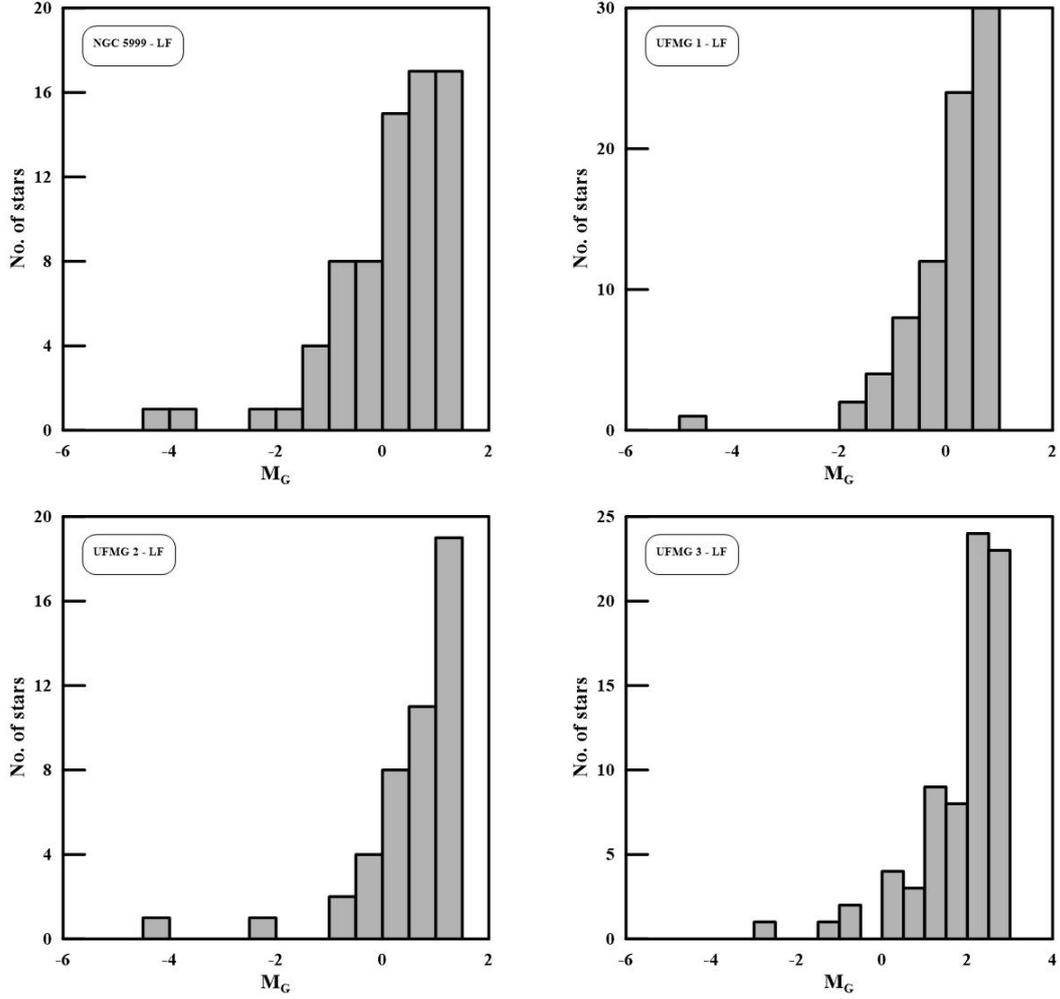

**Fig. 5:** The LFs of NGC 5999 with recent discovered UFMGs open clusters.

On the other hand, and based on the mass-luminosity relation MLR, we turn now to estimate MF of these open clusters, which indicates the relative number of stars formed per unit mass, then we have the initial mass function IMF with power law as follows;

$$\frac{dN}{dM} \propto M^{-\Gamma}. \qquad (40)$$

Where dN/dM represents the number of stars in the mass interval (M : M + dM), and $\Gamma$ is a dimensionless exponent equal to 2.35 according to original Salpeter value (Salpeter 1955).

The adopted isochrones with Marigo et al. (2017) with certain Z and log ($T_{age}$) (yr) in $M_G$ magnitude could construct as a second-order polynomial of MLR in the form:

$$M/M_\odot = a_o + a_1 M_G + a_2 M_G^2. \qquad (41)$$

Where the parameters $(a_o, a_1, a_2)$ and their uncertainties, as well as the mass parameters of the cluster main sequence stars, are given in Table 4.

Scalo (1986, Table IV), calculated the present day mass function PDMF of massive stars (>1 $M_\odot$) and the dimensionless exponent is 4.37 in the region of mass $0 \leq \log (M/M_\odot) \leq 0.54$ (Chabrier 2003). Fig. 6 gives the MFs in logarithmic scale for stars massive than one Solar mass. The power index ($\Gamma$) of PDMF for these open clusters are listed in Table 4; NGC 5999, UFMG 1, and UFMG 3 are in good agreement with that given by Scalo (1986), while that for UFMG 2 is about 0.75 of the Scalo's value. The steep slopes indicate that the numbers of low stars are greater than the high mass ones.

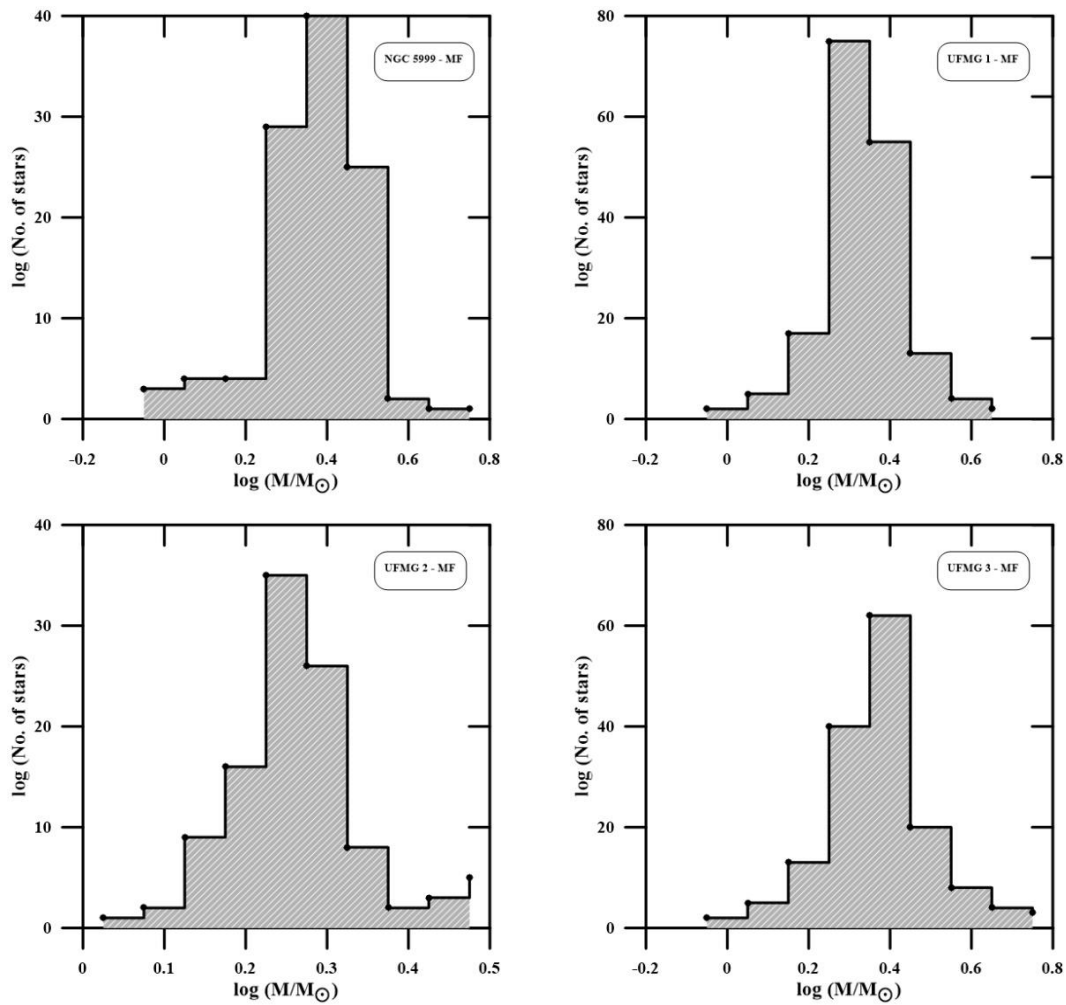

**Fig. 6:** The MFs of NGC 5999 with recent discovered UFMGs open clusters.

## 5. Dynamical relaxation time

- *Tidal radius $r_t$*

In the Galactic disc the sphere of influence of a Gravitational body given by

$$x_L = \left(\frac{GM_C}{4A(A-B)}\right)^{\frac{1}{3}} = \left(\frac{GM_C}{4\Omega_0^2 - \kappa^2}\right)^{\frac{1}{3}}. \quad (42)$$

where $x_L$ is the distance of the Lagrangian points from the center, $G = 4.3 \times 10^{-3}$ pc/M$_\odot$ (km s$^{-1}$)$^2$ is the Gravitational constant, A and B are Oort's constants, $\Omega_0$ the angular velocity and $\kappa$ the epicyclic frequency at the position of the Sun (Röser et al. 2011 and Röser & Schilbach 2019). Here we use A = 15.3 km s$^{-1}$ kpc$^{-1}$ and B = −11.9 km s$^{-1}$ kpc$^{-1}$ from Bovy (2017). The distance $x_L$ is often referred to as the tidal radius $r_t$ of a cluster, i.e. separates, in general, stars Gravitationally bound to a cluster from those that are unbound. Equation (42) including the parameter $M_C$, is the total cumulative mass inside a distance $x_L$ from the center, which can be estimated by integrating the stellar masses over the observed luminosity range; in such way, we refer the counts of $M_G < 0$ as the number of non-MS stars (i.e. $N_{non-MS}$) like Ismail et al. (2015) assuming that those average mass $\langle M/M_\odot \rangle$ is about 1.5 times for that considered MS stars. Then the total mass $M_C$ having the following form:

$$M_C = \left\langle \frac{M}{M_\odot} \right\rangle \left[ N_{MS} + \frac{3}{2} N_{non-MS} \right]. \quad (43)$$

Table 4 listed the total number of MS, non-MS, total mass (with uncertainties), and tidal radius (pc).

- *Processing time*

Open clusters reach a Maxwellian stability equilibrium due to forces of contraction and/or destruction with time called a relaxation time $T_{relax}$ (Maciejewski & Niedzielski, 2007). During that time low mass stars possess the largest random velocity, occupying a larger volume than the high mass does (Mathieu & Latham, 1986). On the other hand, the time needed for the central part of the cluster to be relaxed may be called the dynamical crossing time $T_{cross}$, Lada & Lada (2003) shows that $T_{cross}$ in open clusters is of order $10^6$ years. Mathematically $T_{relax}$ and $T_{cross}$ have the following forms:

$$T_{relax} = \frac{N}{8\ln N} T_{cross}, \quad (44)$$

and

$$T_{cross} = \frac{D}{\sigma}. \tag{45}$$

where N is the number of member stars, σ is the dispersion velocity (Binney & Tremaine 1998), and D is the cluster diameter (Maciejewski & Niedzielski 2007), which corresponds to the limiting radius listed in Table 1, and if the virial theorem is applied, i.e. $\sigma_v^2 \simeq GM_C/r_{lim}$ (Elsanhoury & Amin 2019) one would obtain $T_{cross} \approx 20.00 \pm 4.47$; $27 \pm 5.20$; $25.00 \pm 5.00$; and $25.00 \pm 4.97$ Myr for NGC 5999, UFMG 1, UFMG 2, and UFMG 3 open clusters, respectively.

Adams & Myers (2001) refer the time needed to eject all members from internal stellar encounters as the evaporation time $\tau_{ev}$ ($\gtrsim 10^8$ years) and for a stellar system in virial equilibrium ($\tau_{ev} \sim 10^2\, T_{relax}$). Pinfield et al. (1998) refer that, during $\tau_{ev}$ about 63% of the stars gain enough energy to pass beyond the tidal radius $r_t$, and 'evaporate' off the cluster.

These time-scales (i.e. $\tau_{ev} \gg T_{relax} \sim T_{cross}$) for open clusters, depending on the size of the cluster, the typical mass of the cluster stars, and the number of stars in the cluster.

Finally, the dynamical evolution parameter $\tau = \log (T_{age})/T_{relax}$ describes the dynamical state of the cluster, i.e. if the cluster age founded greater than its relaxation time ($\tau \gg 1$) the clusters may be considered as relaxed one and vice versa. On this context, the first three clusters could classify as dynamically relaxed, while the fourth one is non-relaxed.

**Table 4:** The MFs and dynamical parameters of NGC 5999, UFMG 1, UFMG 2, and UFMG 3 open clusters.

| Parameter | NGC 5999 | UFMG 1 | UFMG 2 | UFMG 3 | Ref. |
|---|---|---|---|---|---|
| $a_o$ | 2.563 ± 0.008 | 2.267 ± 0.005 | 2.043 ± 0.002 | 2.881 ± 0.007 | Present work |
| $a_1$ | -0.451 ± 0.004 | -0.3501 ± 0.003 | -0.2211 ± 0.002 | -0.5678 ± 0.005 | Present work |
| $a_2$ | 0.0276 ± 0.0004 | 0.0206 ± 0.0004 | 0.0047 ± 0.0001 | 0.0386 ± 0.0002 | Present work |
| Γ | -4.17 ± 0.006 | -4.40 ± 0.002 | -3.33 ± 0.002 | -4.83 ± 0.004 | Present work |
| $N_{MS}$ | 83 | 141 | 90 | 134 | Present work |
| $N_{non-MS}$ | 24 | 27 | 8 | 20 | Present work |
| $\langle M/M_{sun}\rangle$ ($M_\odot$) | 2.017 | 1.810 | 1.670 | 2.011 | Present work |
| $M_C$ ($M_\odot$) | 240 ± 16 | 329 ± 18 | 170 ± 13 | 330 ± 18 | Present work |
| $r_t$ (pc) | 8.53 ± 2.92 | 9.48 ± 3.08 | 7.60 ± 2.76 | 9.48 ± 3.10 | Present work |
| $T_{cross}$ (Myr) | 20.00 ± 4.47 | 27.00 ± 5.20 | 25.00 ± 5.00 | 25.00 ± 4.97 | Present work |
| $T_{relax}$ (Myr) | 57.00 ± 7.55 | 111 ± 10.54 | 67.00 ± 8.20 | 95.54 ± 9.77 | Present work |
| $\tau_{ev}$ (Myr) | 5700 ± 75 | 11100 ± 105.36 | 6700 ± 82.00 | 9554 ± 98.00 | Present work |
| $\tau$ | 5.55 ± 2.36 | 7.16 ± 2.68 | 22 ± 4.60 | 1.05 ± 0.12 | Present work |

## 6. Conclusion

Recent data of Gaia Collaboration with accurate measurements of astrometric parameters and radial velocities motivated us to carry out the present work; in this purpose, a computational routine using the *Mathematica* software has been developed. Therefore, we have determined for the first time, various kinematical and dynamical structure of the newly discovered UFMGs clusters in the vicinity of NGC 5999. The conclusion reaches as follows:

- For our member estimation (i.e. N = 107, 168, 98, and 154) for NGC 5999 and UFMGs clusters respectively, we calculated the coordinate positions (apex) by AD-diagram method, showing that ($102^o.40 \pm 1.02$ & $-4^o.60 \pm 0.47$; NGC 5999), ($96^o.69 \pm 1.10$ & $-0^o.58 \pm 0.045$; UFMG 1), ($97^o.47 \pm 1.09$ & $1^o.56 \pm 0.051$; UFMG 2), and ($98^o.65 \pm 1.12$ & $-0^o.26 \pm 0.060$; UFMG 3), also different kinematical parameters have been computed like distances r (kpc), i.e. $2.16 \pm 0.046$; NGC 5999, $1.77 \pm 0.042$; UFMG 1, $1.96 \pm 0.044$; UFMG 2, and $1.55 \pm 0.039$; UFMG 3 which are in good argument as compared with Ferreira et al. (2016), space velocities $\sqrt{\left(\overline{U}^2 + \overline{V}^2 + \overline{W}^2\right)}$, clusters centers ($x_c$, $y_c$, $z_c$), projected distances ($X_\odot$, $Y_\odot$, $Z_\odot$), and the Solar elements ($S_\odot$, $l_A$, $b_A$).

- According to the mass-luminosity relation MLR, we have computed the total mass $M_C$ of these objects by adopted isochrones with Marigo et al. (2017) for certain metallicities (Z) and log ($T_{age}$) (yr) in $M_G$ magnitude region. Total mass including both $N_{MS}$ and $N_{non-MS}$ stars, in such a way, $M_C$ ($M_\odot$) takes the values, $240 \pm 16$; NGC 5999, $329 \pm 18$; UFMG 1, $170 \pm 13$; UFMG 2, and $330 \pm 18$; UFMG 3.

- For our targets, we have computed their tidal radii $r_t$ (pc) as a function of total mass $M_C$ (i.e. $8.53 \pm 2.92$; NGC 5999, $9.48 \pm 3.08$; UFMG 1, $7.60 \pm 2.76$; UFMG 2, and $9.48 \pm 3.10$; UFMG 3).

- Our computed evaporation time $\tau_{ev}$ seems $\gg T_{cross}$ and $T_{relax}$, which consistent with other calculation of open clusters evaporation time. Finally, NGC 5999, UFMG 1, and UFMG 2 could classify as relaxed clusters, while UFMG 3 is non-relaxed one.

*Acknowledgments*

This research has made use of the VizieR catalog access tool, CDS, Strasbourg, France. The preparation of this work has made extensive use of NASA's Astrophysics Data System Bibliographic Services. This work has made use of data from the European Space Agency (ESA) mission Gaia https://www.cosmos.esa.int/gaia, the data from which were processed by the Gaia Data Processing and Analysis Consortium (DPAC). Funding for the DPAC has been provided by national institutions, in particular, the institutions participating in the Gaia Multi-Lateral Agreement (MLA). The Gaia archive website is https://archives.esac.esa.int/gaia.**References**

1. Adams, F. C. and Myers, P. 2001, *ApJ*, 533, 744.
2. Bica, E., Alloin, D. 1986, *A&A*, 162, 21.
3. Bica, E., Alloin, D. 1987, *A&A*, 186, 49.
4. Bovy, J. 2017, *MNRAS*, 468, L63.
5. Cantat-Gaudin, T., Jordi, C. et al. 2018, *A&A*, 618, 93C.
6. Chabrier, G. 2003, *Publ. Astron. Soc. Pac.*, 115, 763.
7. Chumak, Y. O., Rastorguev, A. S. 2006, *Astronomy Letters*, 32, 446.
8. Chupina, N. V., Reva, V. G., Vereshchagin, S. V. 2001, *A&A*, 371, 115.
9. Chupina, N. V., Reva, V. G., Vereshchagin, S. V. 2006, *A&A*, 451, 909.
10. Dias, W. S., Alessi, B. S., Moitinho, A., Lépine, J. R. D. 2002, *A&A*, 389, 871.
11. Eggen, O. J. 1984, *AJ*, 89, 1350.
12. Elsanhoury, W. H, Sharaf, M. A, Nouh, M. I., and Saad, A. S. 2013, *OAJ*, 6, 1.
13. Elsanhoury, W. H., Nouh, M. I. and Abdel-Rahman, H. I. 2015, *Rev. Mex. Astron. Astrofís.,* 51, 197.
14. Elsanhoury, W. H., Haroon, A. A., Chupina, N. V., Vereshchagin, S. V., Sariya Devesh, P., Yadav, R. K. S., Jiang, I.-G. 2016, *New Astron.*, 49, 32.
15. Elsanhoury, W. H., Postnikova, E. S., Chupina, N. V., Vereshchagin, S. V., Devesh, P. Sariya, Yadav, R. K. S. and Jiang, I. 2018, *ApSS*, 363, Issue 3, 58.
16. Elsanhoury, W. H. and Nouh, M. I. 2019, 72, 19.
17. Elsanhoury, W. H. and Amin, M. Y. 2019, *Serb. Astron. J.*, in press.
18. Evans, D. W., et al. 2018, *A&A*, 616, A4.
19. Ferreira, F. A., Santos, J. F. C., et al. 2019, *MNRAS*, 483, 4.
20. Francis, C. and Anderson, E. 2014, *MNRAS*, 441, 1105.
21. Fürnkranz, V., Meingast, S., Alves J. 2019, *arXiv:1902.07216v2*.
22. Gaia Collaboration et al. 2018, *A&A*, 616, A1.
23. Gunn, J. E., Griffin, R. F., Griffin, R. F. M. and Zimmerman, B. A. 1988, *AJ*, 97, 198.
24. Hanson, R. B. 1975, *AJ*, 80, 379.